\documentclass[preprint,5p]{elsarticle}

\usepackage{amsmath,amssymb,bm}
\usepackage{lineno,hyperref}

\journal{Physics Letter B}
\begin{document}

\begin{frontmatter}

\title{Nuclear masses in extended kernel ridge regression with odd-even effects}

\author[PKU]{X. H. Wu}
\author[PKU]{L. H. Guo}
\author[PKU]{P. W. Zhao.\corref{cor1}}
\ead{pwzhao@pku.edu.cn}

%\date{\today}

\cortext[cor1]{Corresponding Author}
\address[PKU]{State Key Laboratory of Nuclear Physics and Technology, School of Physics, Peking University, Beijing 100871, China}

\begin{abstract}
  The kernel ridge regression (KRR) approach is extended to include the odd-even effects in nuclear mass predictions by remodulating the kernel function without introducing new weight parameters and inputs in the training network.
  By taking the WS4 mass model as an example, the mass for each nucleus in the nuclear chart is predicted with the extended KRR network, which is trained with the mass model residuals, i.e., deviations between experimental and calculated masses, of other nuclei with known masses.
  The resultant root-mean-square mass deviation from the available experimental data for the 2353 nuclei with $Z\ge8$ and $N\ge8$ can be reduced to 128 keV, which provides the most precise mass model from machine learning approaches so far.
  Moreover, the extended KRR approach can avoid the risk of worsening the mass predictions for nuclei at large extrapolation distances, and meanwhile, it provides a smooth extrapolation behavior with respect to the odd and even extrapolation distances.
\end{abstract}

%\pacs{21.10.Re, 23.20.Lv, 27.50.+e}
%\maketitle
\end{frontmatter}

%\linenumbers
%======================Introduction========================%
\section{Introduction}

Nuclear mass is of fundamental importance not only for various aspects of nuclear physics~\cite{Lunney2003Rev.Mod.Phys.1021}, but also for astrophysics~\cite{Burbidge1957Rev.Mod.Phys.547, Mumpower2016Prog.Part.Nucl.Phys.86}.
It can be used to extract a lot of nuclear structure information, e.g., nuclear deformation~\cite{Hager2006Phys.Rev.Lett.42504, Roubin2017Phys.Rev.C14310}, shell structure \cite{Ramirez2012Science1207, Wienholtz2013Nature346}, and effective interactions~\cite{AlexBrown1998Phys.Rev.C220, Lalazissis2005Phys.Rev.C24312, Zhao2010Phys.Rev.C54319}.
It is also a key nuclear physics input in understanding the energy production in stars \cite{Bethe1939Phys.Rev.434} and the origin of elements in the universe \cite{Burbidge1957Rev.Mod.Phys.547} by determining the reaction energies of all involved nuclear reactions.

Great achievements in nuclear mass measurements have been made during recent decades with the development of radioactive ion beam (RIB) facilities, and about 2500 nuclear masses have been measured to date~\cite{Wang2012Chin.Phys.C1603, Wang2017Chin.Phys.C30003}.
Nevertheless, the masses of a large number of neutron-rich nuclei involved in the $r$-process remain unknown from experiments and cannot be measured even with the next-generation RIB facilities.
The local mass relations, such as the isobaric multiplet mass equation (IMME)~\cite{Ormand1997Phys.Rev.C2407}, the Garvey-Kelson (GK) relations~\cite{Barea2008Phys.Rev.C41304}, and the residual proton-neutron interactions~\cite{Fu2011Phys.Rev.C34311}, can be used to predict the masses of nuclei very close to the experimentally known region, but they are not sufficient for the demands of $r$-process simulations.
Therefore, theoretical predictions for nuclear masses are imperative at the present time.
It can be traced back to the von Weizs\"acker mass formula based on the famous liquid drop model (LDM)~\cite{Weizsaecker1935ZeitschriftfurPhysik431}.
Tremendous efforts have been made in pursuing different possible extensions of the LDM, which are known as the macroscopic-microscopic models, such as the finite-range droplet model (FRDM)~\cite{Moeller2016Atom.DataNucl.DataTables1} and the Weizs\"acker-Skyrme (WS) model~\cite{Wang2014Phys.Lett.B215}. The microscopic mass models based on the nonrelativistic and relativistic density functional theories (DFTs) have also been developed (see e.g., Refs.~\cite{Goriely2013Phys.Rev.C61302,Geng2005Prog.Theor.Phys.785,Erler2012nat509,Afanasjev2013PhysicsLettersB680,Lu2015Phys.Rev.C27304,Xia2018Atom.DataNucl.DataTables1,Yang2020ChinesePhysicsC34102} and references therein).
They are usually believed to have a better reliability of extrapolation~\cite{Zhao2012Phys.Rev.C64324}, although their precisions of predicting experimentally known masses are currently poorer than the macroscopic-microscopic models.

The root-mean-square (rms) deviations between theoretical mass models and the available experimental data~\cite{Wang2012Chin.Phys.C1603,Wang2017Chin.Phys.C30003} range from about 3 MeV for the BW model \cite{Kirson2008Nucl.Phys.A29} to about 300 keV for the WS ones \cite{Wang2014Phys.Lett.B215}, which are still not enough for accurate studies of exotic nuclear structure and astrophysical nucleosynthesis.
In particular, for neutron-rich nuclei far away from the experimentally known region, the differences among the predictions of different mass models can be as large as several tens MeV.
In recent years, machine learning approaches have been employed to further improve the accuracies of nuclear models, such as the radial basis function (RBF) approach~\cite{Wang2011Phys.Rev.C51303, Niu2013Phys.Rev.C24325, Zheng2014Phys.Rev.C14303, Niu2016Phys.Rev.C54315, Ma2017Phys.Rev.C24302, Niu2018Sci.Bull.759}, the Bayesian neural network (BNN) approach~\cite{Utama2016Phys.Rev.C14311, Utama2017Phys.Rev.C44308, Niu2018Phys.Lett.B48, Neufcourt2018Phys.Rev.C34318}, and the kernel ridge regression (KRR) approach~\cite{Wu2020Phys.Rev.C051301}.
By training the machine learning network with the mass model residuals, i.e., deviations between experimental and calculated masses, machine learning approaches can reduce the corresponding rms deviations to about $200$ keV.
However, the RBF and the BNN approaches predict quite different masses for nuclei far away from the region with known masses \cite{Niu2019Phys.Rev.C54311}.
This indicates that the extrapolation abilities of these two machine learning tools have not been properly understood yet.
In contrast, the KRR approach with the Gaussian kernel can automatically identify the limit of the extrapolation distance and avoid the risk of worsening the mass descriptions for nuclei at large extrapolation distances~\cite{Wu2020Phys.Rev.C051301}.

The rms deviations can be further reduced with more effects being taken into account, e.g., the odd-even effects, pairing effects, and shell effects.
The odd-even effects are included in the RBF approach by building and training additional networks~\cite{Niu2016Phys.Rev.C54315}, while the pairing and shell effects are included in the BNN approach~\cite{Niu2018Phys.Lett.B48} by including additional inputs in the network.
The numbers of weight parameters are significantly increased in comparison with the original networks in both approaches.
The accuracies of the mass predictions for known nuclei are further improved, but the RBF and the BNN approaches still predict quite different masses for nuclei far away from the known region~\cite{Niu2019Phys.Rev.C54311}.

In the present work, the KRR approach is extended to include the odd-even effects for nuclear mass predictions.
The odd-even effects are included only by remodulating the KRR kernel function. Therefore, the number of the weight parameters is not increased, and the inputs of the network remain in the present extended KRR approach.
The hyperparameters in the extended KRR network are optimized by careful cross-validations.
The performance and reliability of the extrapolated mass predictions are analyzed in detail.

%==================Theoretical framework===================%
\section{Theoretical framework}

The KRR approach is a powerful machine learning approach for nonlinear regression and has been successfully applied in nuclear mass predictions~\cite{Wu2020Phys.Rev.C051301}. In order to include the odd-even effects, the KRR function is extended to be
\begin{equation}\label{KRR_function}
  S(\bm{x}_j) = \sum_{i=1}^{m}K(\bm{x}_j,\bm{x}_i)\alpha_i + \sum_{i=1}^{m}K_{\rm oe}(\bm{x}_j,\bm{x}_i)\beta_i,
\end{equation}
where $\bm{x}_i \equiv (Z_i, N_i)$ are locations of nuclei in the nuclear chart, $m$ is the number of training nuclei, $\alpha_i$ and $\beta_i$ are weights to be determined, $K(\bm{x}_j,\bm{x}_i)$ and $K_{\rm oe}(\bm{x}_j,\bm{x}_i)$ are kernels, which measure the similarity between nuclei.
The closer two nuclei on the nuclear chart, the larger similarity they have. This is measured by the Gaussian kernel
\begin{equation}\label{Gaussian_kernel}
  K(\bm{x}_j,\bm{x}_i) = \exp(-||\bm{x}_i-\bm{x}_j||^2/2\sigma^2),
\end{equation}
where $||\bm{x}_i-\bm{x}_j|| = \sqrt{(Z_i-Z_j)^2+(N_i-N_j)^2}$ stands for the distance of two nuclei.
The kernel $K_{\rm oe}(\bm{x}_j,\bm{x}_i)$ is introduced to enhance the correlations between nuclei that have the same parity of proton and neutron numbers, which reads
\begin{equation}\label{Gaussian_kernel}
  K_{\rm oe}(\bm{x}_j,\bm{x}_i) = \delta_{\rm oe}(\bm{x}_j,\bm{x}_i)\exp(-||\bm{x}_i-\bm{x}_j||^2/2{\sigma_{\rm oe}}^2).
\end{equation}
Here, $\delta_{\rm oe}(\bm{x}_j,\bm{x}_i) = 1$ for two nuclei have the same parity of proton and neutron numbers, otherwise $\delta_{\rm oe}(\bm{x}_j,\bm{x}_i) = 0$.
The Gaussian widths $\sigma$ and $\sigma_{\rm oe}$ measure the length scale on the distance that the kernels affect.
The weights $\alpha_i$ and $\beta_i$ are determined by minimizing the loss function defined as
\begin{equation}\label{Loss_function}
  L(\bm{\alpha},\bm{\beta}) = \sum_{i=1}^{m} [S(\bm{x}_i) - y(\bm{x}_i)]^2 + \lambda \bm{\alpha}^T\bm{K}\bm{\alpha} + \lambda_{\rm oe}\bm{\beta}^T\bm{K}_{\rm oe}\bm{\beta}.
\end{equation}
The first term is the variance between the data $y(\bm{x}_i)$ and the KRR prediction $S(\bm{x}_i)$, and the second and third terms are regularizers that penalize large weights to reduce the risk of overfitting. The hyperparameters $\lambda$ and $\lambda_{\rm oe}$ determine the regularization strength. Minimizing Eq.~\eqref{Loss_function}, we can obtain
\begin{align}
  & \bm{\beta} = \frac{\lambda}{\lambda_{\rm oe}}\bm{\alpha} , \label{eqbeta} \\
  & \bm{\alpha} = \left( \bm{K} + \frac{\lambda}{\lambda_{\rm oe}}\bm{K}_{\rm oe} + \lambda\bm{I}  \right)^{-1}\bm{y}. \label{eqalpha}
\end{align}
With Eq.~\eqref{eqbeta}, the extended KRR function \eqref{KRR_function} can be rewritten as a standard KRR function
\begin{equation}\label{KRR_function_new}
  S(\bm{x}_j) = \sum_{i=1}^{m} K'(\bm{x}_j,\bm{x}_i) \alpha_i,
\end{equation}
with the remodulating kernel
\begin{equation}\label{remodu_kernel}
  K'(\bm{x}_j,\bm{x}_i)=K(\bm{x}_j,\bm{x}_i) + \frac{\lambda}{\lambda_{\rm oe}}\bm{K}_{\rm oe}(\bm{x}_j,\bm{x}_i).
\end{equation}
Here, the weights $\alpha_i$ are determined by Eq.~\eqref{eqalpha}.
Note that the number of weight parameters in the present extended KRR approach is the same as the original KRR due to the relation  between $\bm{\alpha}$ and $\bm{\beta}$ in Eq.~\eqref{eqbeta}.

%==================Numerical details=======================%
\section{Numerical details}

The extended KRR function \eqref{KRR_function_new} is trained to reconstruct the mass residuals, i.e., the deviations $\Delta M (N,Z) = M_{\rm exp}(N,Z) - M_{\rm th}(N,Z)$ between the experimental data $M_{\rm exp}$ and predictions $M_{\rm th}$ for a given mass model.
Once the weights $\alpha_i$ are obtained, the extended KRR function $S(N,Z)$ can be calculated with Eq.~\eqref{KRR_function_new} for every nucleus $(N,Z)$.
The predicted mass for a nucleus $(N,Z)$ is, thus, given by $M_{\rm EKRR}(N,Z) = M_{\rm th}(N,Z) + S(N,Z)$.

In the present work, the 2353 experimental data of masses $M_{\rm exp}$ for nuclei with $Z\geq8$ and $N\geq8$ are taken from AME2012~\cite{Wang2012Chin.Phys.C1603}.
The theoretical masses $M_{\rm th}$ are taken from the mass table WS4~\cite{Wang2014Phys.Lett.B215}, which is one of the most accurate nuclear mass tables.

The leave-one-out cross-validation is adopted to determine the hyperparameters $(\sigma,\lambda,\sigma_{\rm oe},\lambda_{\rm oe})$.
With a given set of hyperparameters $(\sigma,\lambda,\sigma_{\rm oe},\lambda_{\rm oe})$, the mass prediction for each of the 2353 nuclei can be given with the extended KRR network trained on all other 2352 nuclei.
The optimized hyperparameters $(\sigma=1.25,\lambda=0.05,\sigma_{\rm oe}=2.65,\lambda_{\rm oe}=0.15)$ are obtained when the rms deviation between experimental and predicted masses of the 2353 nuclei attains its minimum.

%================Results and discussion====================%
\section{Results and discussion}

\begin{figure}[!htbp]
  \centering
  \includegraphics[width=0.45\textwidth]{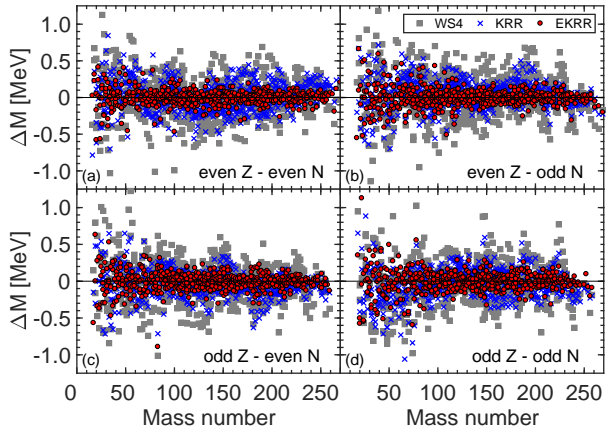}
  \caption{(Color online). Mass differences between the experimental data and the predictions of the WS4, the KRR, and the extended KRR models for the (a) even-even, (b) even-odd, (c) odd-even, and (d) odd-odd nuclei.
  }\label{fig1}
\end{figure}

The mass predictions $M_{\rm EKRR}$ for every nuclei can be calculated by the extended KRR approach with the determined hyperparameters, in which the network weights $\alpha_i$ are trained on the set consisting of all other nuclei with experimentally known masses.

The mass differences between the extended KRR predictions and the experimental data are shown in Fig.~\ref{fig1} for the even-even (e-e), even-odd (e-o), odd-even (o-e), and odd-odd (o-o) nuclei, in comparison with the ones of the WS4~\cite{Wang2014Phys.Lett.B215} and the KRR predictions~\cite{Wu2020Phys.Rev.C051301}.
It can be clearly seen that the predictive power of the WS4 mass model is further improved by the extended KRR approach in comparison with the
KRR approach.
The significant improvement of the extended KRR approach is mainly due to the consideration of the odd-even effects, which eliminates the staggering behavior of mass deviations with respect to the even and odd numbers of nucleons in the KRR approach~\cite{Wu2020Phys.Rev.C051301}.
Quantitatively, the rms deviation $\Delta_{\rm rms}=298$~keV of 2353 nuclei for the WS4 mass model, is reduced to $199$~keV by the KRR approach, and is further reduced to $128$~keV by the extended KRR approach with the odd-even effects.
This indeed provides, so far, the most precise mass model from machine learning approaches.

\begin{figure}[!htbp]
  \centering
  \includegraphics[width=0.45\textwidth]{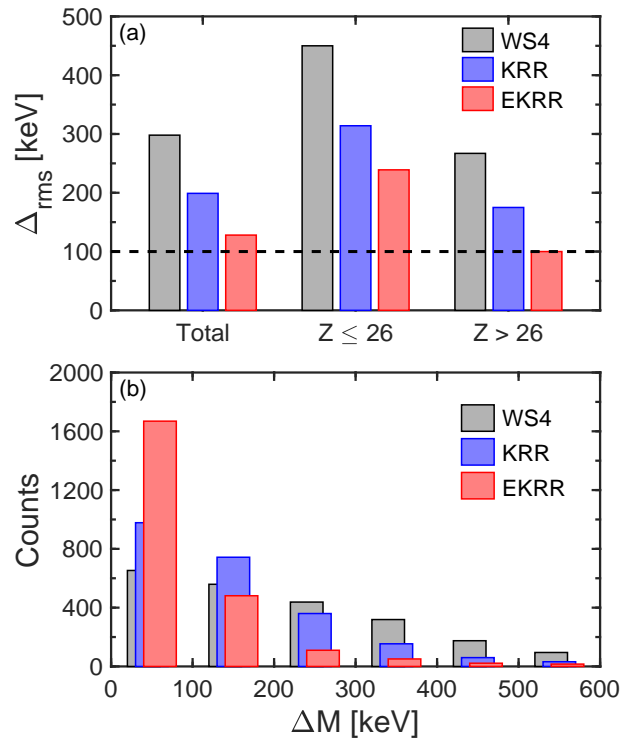}
  \caption{(Color online). (a) The rms deviations $\Delta_{\rm rms}$ between the experimental data and the predictions of the WS4, the KRR, and the extended KRR models for $Z\leq 26$, $Z>26$ nuclei, and the whole set of nuclei. The dashed line is drawn for guiding eyes at the rms deviation of 100 keV.
  (b) The number of nuclei with the corresponding mass deviations from the data $\Delta M$ locating in various slots, such as $ 0 < \Delta M \le 100 $~keV, $ 100 < \Delta M \le 200 $~keV, etc.
  }\label{fig2}
\end{figure}

The mass deviations between the data and the predictions are relatively larger for light nuclei.
This may be due to the fact that there are large individual differences among nuclear masses of light nuclei, which are difficult to be reproduced by global mass models.
On the other hand, for the applications of nuclear masses in the $r$-process simulations, the masses of nuclei heavier than iron are important.
In Fig.~\ref{fig2}(a), the rms deviations $\Delta_{\rm rms}$ between the experimental data and the predictions of the WS4, the KRR, and the extended KRR models for light ($Z \leq 26$), heavy ($Z > 26$), and the whole set of nuclei are shown.
The extended KRR approach improves the descriptions of masses for both light and heavy nuclei significantly.
Moreover, for the heavy nuclei with $Z > 26$, the rms deviation between the experimental data and
the predictions of the extended KRR model is as small as $100$~keV, which is reaching the chaos-related unpredictability limit for the calculation of nuclear masses~\cite{Barea2005Phys.Rev.Lett.102501, Niu2018Sci.Bull.759}.

In Fig.~\ref{fig2}(b), it is depicted that the number of nuclei with the corresponding mass deviations from the data $\Delta M$ locating in various slots, such as $0<\Delta M \leq 100$~keV, $100<\Delta M \leq 200$~keV, etc.
This gives a detailed analyse for the distributions of the mass deviations.
For the KRR model, the mass deviations from data are smaller than 200 keV for most nuclei, and the number of these nuclei is 1721, which is about 73\% of the whole nuclei set.
For the extended KRR model, however, the mass deviations from data are smaller than 100 keV for more than 70\% of the nuclei, and they are smaller than 200 keV for more than 90\% of the nuclei.
This means that there are only $200$ nuclei with mass deviations larger than 200 keV, and most of them are light nuclei.
To improve the descriptions of masses for these nuclei would be a challenging task for the future.

\begin{figure}[!htbp]
  \centering
  \includegraphics[width=0.45\textwidth]{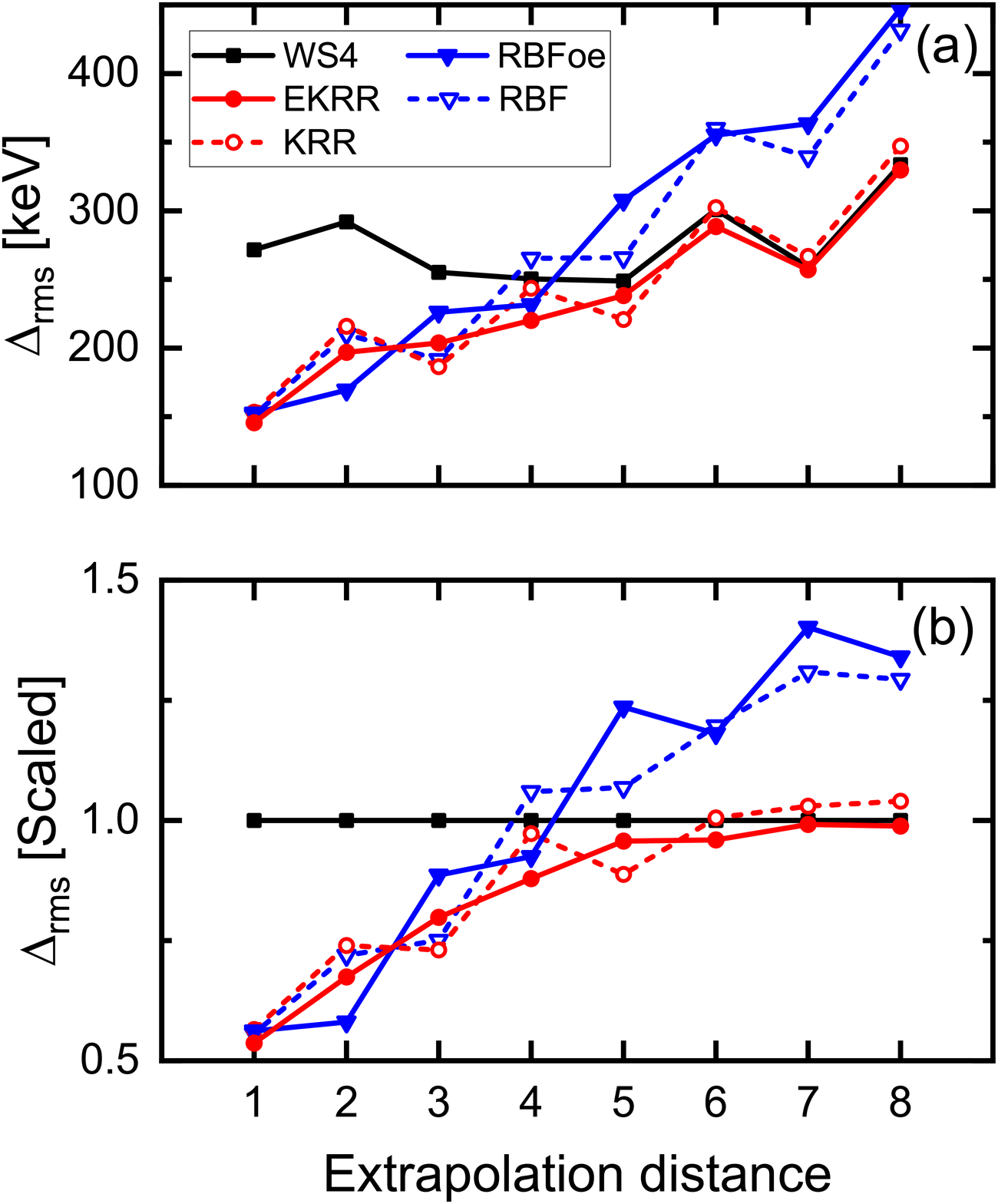}
  \caption{(Color online). Comparison of the extrapolation power of the extended KRR, the KRR, the RBF and the RBF approach with odd-even effects (RBFoe)~\cite{Niu2016Phys.Rev.C54315} for eight test sets with different extrapolation distances (see text for details). (a) The rms deviations $\Delta_{\rm rms}$ of the calculated masses from the WS4 mass model (filled squares), the extended KRR (filled circles), the KRR (empty circles), the RBF (empty triangles), and the RBFoe extrapolations (filled triangles)  with respect to the experimental masses. (b) For a clearer comparison, the rms deviations $\Delta_{\rm rms}$ are scaled to the values for the WS4 mass model.
  }\label{fig3}
\end{figure}

To examine the extrapolation power of the extended KRR approach for neutron-rich nuclei, similar to Ref.~\cite{Wu2020Phys.Rev.C051301},
for each isotopic chain with $Z>26$, the eight most neutron-rich nuclei are removed out from the training set, and they are classified into eight test sets respectively, corresponding to the different extrapolation distances from the remain training set in the neutron direction.

In Fig.~\ref{fig3}(a), the rms deviations $\Delta_{\rm rms}$ of the calculated masses for the eight test sets from the WS4 mass model, the extended KRR extrapolations, the KRR ones, the RBF ones, and the RBF ones with odd-even effects (RBFoe)~\cite{Niu2016Phys.Rev.C54315} with respect to the experimental masses are shown as functions of the extrapolation distance.
An even more clear comparison is shown in Fig.~\ref{fig3}(b), where the rms deviations $\Delta_{\rm rms}$ are scaled to the corresponding ones for the WS4 mass model.
One can see that all the four approaches improve the mass descriptions of nuclei with the extrapolation distances smaller than $4$.
The corresponding rms deviations are reduced by up to $\approx 40\%$ in comparison with the WS4 mass model.
However, one can see obvious distinct features between the KRR type of approaches and the RBF type of approaches at extrapolation distances larger than $4$.
The rms deviations of the RBF and the RBFoe extrapolations are larger than the ones of the WS4 mass model, and they increase rapidly with the extrapolation distance.
For the KRR and the extended KRR extrapolations, the rms deviations increase slowly with the extrapolation distance and are similar or even smaller than the WS4 ones at large extrapolation distances.
The distinct features are due to the different behaviors of the Gaussian kernel used in the KRR type of approaches and the Linear kernel used in the RBF type of approaches, which have been discussed in detail in Ref.~\cite{Wu2020Phys.Rev.C051301}.

Obvious odd-even staggerings along the extrapolation distance can be seen in the rms deviations for the RBF and KRR predictions, where the odd-even effects have not been taken into account explicitly.
By considering the odd-even effects with the RBFoe approach, however, the odd-even staggerings remain, and even with an opposite phase with respect to the RBF approach.
This means that the RBFoe approach may overestimate the odd-even effects.
On the contrary, for the extended KRR approach, as shown in Fig.~\ref{fig3}(b), the rms deviations vary smoothly with the extrapolation distance.
This indicates that the odd-even effects have been well handled in the extended KRR approach.
Moreover, with the increase of the extrapolation distance, the extended KRR approach converged to the original WS4 mass model.
This keeps the key merit of the KRR approach for mass extrapolations, i.e., it avoids worsening the mass descriptions for nuclei at large extrapolation distances.

%===============SUMMARY ==============
\section{Summary}

In summary, the kernel ridge regression approach has been extended to include the odd-even effects in nuclear mass predictions by remodulating the kernel function.
The obtained extended KRR approach does not introduce new weight parameters and inputs in the training network.
The resulting rms mass deviation from the experimental data is reduced from $298$~keV for the WS4 mass model to $128$~keV for the extended KRR approach, which provides the most precise mass model from machine learning approaches so far.
For nuclei heavier than iron, in particular, the rms deviation is as small as 100 keV, which is reaching the chaos-related unpredictability limit for the calculation of nuclear masses.
Comparative study has been carried out to examine the extrapolation power of the extended KRR approach to the neutron-rich nuclei region.
It reveals that the extended KRR approach can avoid the risk of worsening the mass predictions for nuclei at large extrapolation distances, and meanwhile, it provides a smooth extrapolation behavior with respect to the odd and even extrapolation distances.

%===============Acknowledgments ==============

\section*{Acknowledgments}

This work was partly supported by the National Key R\&D Program of China (Contracts No. 2018YFA0404400 and No. 2017YFE0116700), the National Natural Science Foundation of China (Grants No. 11875075, No. 11935003, No. 11975031 and No. 12070131001).

%\section*{References}

\end{document}